\documentclass[preprintnumbers,amsmath,amssymb,aps,prl]{revtex4}
\usepackage{amsmath}
\usepackage[mathcal]{eucal}
\usepackage{latexsym}
\usepackage{amssymb}
\usepackage{color}
\definecolor{Red}{rgb}{1.,0.,0.}
\definecolor{Blue}{rgb}{0.,0.,1.}

\newcommand*{\eps}{{\rlap{\lower2ex\hbox{$\,\,\tilde{}$}}{\epsilon_{ijk}}}}
\newcommand*{\EPS}{{\rlap{\lower2ex\hbox{$\,\,\tilde{}$}}{\epsilon_{i'j'k'}}}}
\newcommand*{\lmq}{{\rlap{\lower2ex\hbox{$\,\,\tilde{}$}}{\epsilon_{lmq}}}}
\newcommand*{\jmq}{{\rlap{\lower2ex\hbox{$\,\,\tilde{}$}}{\epsilon_{jmq}}}}
\newcommand*{\jql}{{\rlap{\lower2ex\hbox{$\,\,\tilde{}$}}{\epsilon_{jql}}}}
\newcommand*{\jlm}{{\rlap{\lower2ex\hbox{$\,\,\tilde{}$}}{\epsilon_{jlm}}}}
\newcommand*{\imq}{{\rlap{\lower2ex\hbox{$\,\,\tilde{}$}}{\epsilon_{imq}}}}
\newcommand*{\iql}{{\rlap{\lower2ex\hbox{$\,\,\tilde{}$}}{\epsilon_{iql}}}}
\newcommand*{\ilm}{{\rlap{\lower2ex\hbox{$\,\,\tilde{}$}}{\epsilon_{ilm}}}}
\newcommand*{\lmn}{{\rlap{\lower2ex\hbox{$\,\,\tilde{}$}}{\epsilon_{lmn}}}}
\newcommand*{\abc}{{\rlap{\lower2ex\hbox{$\,\,\tilde{}$}}{\epsilon_{abc}}}}
\newcommand*{\N}{{\rlap{\lower2ex\hbox{$\,\,\tilde{}$}}{N}}}
\newcommand{\tN}{{\rlap{\lower2ex\hbox{$\,\,\tilde{}$}}{N}}}
\newcommand*{\tM}{{\rlap{\lower2ex\hbox{$\,\,\tilde{}$}}{M}}}
\newcommand*{\I}{{\rlap{\lower2ex\hbox{$\,\,\tilde{}$}}{e_{i}^{a}}}}
\newcommand*{\J}{{\rlap{\lower2ex\hbox{$\,\,\tilde{}$}}{e_{j}^{a}}}}

\begin{document}
\title{Affine group formulation of the Standard Model coupled to gravity}

\author{Ching-Yi Chou}\email{l2897107@mail.ncku.edu.tw}
\address{Department of Physics, National Cheng Kung University, Taiwan}
\author{Eyo Ita}\email{ita@usna.edu}
\address{Department of Physics, US Naval Academy, Annapolis Maryland}
\author{Chopin Soo}\email{cpsoo@mail.ncku.edu.tw}
\address{Dept. of Physics, National Cheng Kung University, Taiwan}
\input amssym.def
\input amssym.tex



\bigskip

\begin{abstract}
This work demonstrates that a complete description of the
interaction of matter and all forces, gravitational and
non-gravitational,  can in fact be realized within a quantum affine
algebraic framework. Using the affine group formalism, we construct elements of a physical
Hilbert space for full, Lorentzian quantum gravity coupled to the Standard Model in four spacetime dimensions.  
Affine algebraic quantization of gravitation and matter on equal footing implies a fundamental uncertainty relation which is predicated upon a non-vanishing cosmological constant.

\end{abstract}
\maketitle

\section{Introduction}

In this work a new approach to the Wheeler-DeWitt constraint is
presented. This approach will be based upon the affine group
formalism, utilizing unitary irreducible representations of the
group of transformations of the straight line $x\rightarrow{a}x+b$.
Previous results by the present authors reveal a fundamental
structure concealed within the Hamiltonian constraint of General
Relativity (GR), which has provided insight into such an approach.
The background and results needed for this paper are encapsulated in
\cite{CIS}, \cite{LASZLO}, \cite{LASZLO1} and references therein.
In \cite{CIS} it was shown that the quantum Hamiltonian constraint
of vacuum GR with cosmological constant can be written as an affine
algebra involving the commutator of the imaginary part of the
Chern--Simons funtional, $Q$, with the local volume operator $V(x)$,
whose representations led directly to the construction of elements
of a physical Hilbert space for gravity respecting the intital value
constraints. The classical aspects of these results were
independently derived in \cite{LASZLO}, in the metric
representation, from a different context. In \cite{LASZLO} the
Hamiltonian constraint was written as a Poisson bracket of $Q$ with
$V(x)$, which provided the insight of the conformal invariance of
the real part of the Chern--Simons functional, a symmetry which
becomes broken by its imaginary part. The necessary and sufficient
condition for the invariance of this imaginary part with respect to
infinitesimal rescalings of the spacetime metric is just that the
spacetime metric satisfy the vacuum Einstein equations. In
\cite{LASZLO1} the results of \cite{LASZLO} were generalized to
include the Higgs and the Yang--Mills fields, showing that the
Hamiltonian constraint of the coupled theory can also be written as
the Poisson bracket of a certain functional $G$ with the volume
element.  The kinematic constraints were not addressed.

In the present paper we will generalize these results, with the goal
of quantization of the Standard Model coupled to gravity.  One
motivation stems from the observation that purely quantum
gravitational effects presumably, by themselves, may not become
directly observable until the Planck scale, several orders of
magnitude beyond typical energies accessible to existing
accelerators.  A consistent treatment of the gravity coupling with
matter, which quantizes all fields on an equal footing, should
in-principle encode the quantum effects of gravity in this
interaction.  Conceivably, the imprints of gravity from the Planck
scale could be encoded within observations which could be made on matter
fields at everyday energy scales.  This paper will constitute an
attempt to bridge this gap via a proposal designed to respect the
fundamental assumption of background independence in congruity with
the axioms of quantum mechanics.\par \indent We will ultimately
construct elements of a physical Hilbert space $\textbf{H}_{Phys}$,
which respects the initial value constraints (Hamiltonian,
diffeomorphism and Gauss' law constaints), for gravity coupled with
matter fields.  In order to carry out this goal we will need to put
in place the following elements: (i) First, we will generalize the
results of \cite{LASZLO1} to include coupling to Weyl spinor fields.
Concurrently, we will adapt the matter Hamiltonians of
\cite{LASZLO1} to density weight one form to bring them into
congruity with \cite{CIS}. This will be necessary in order to
address the diffeomorphism-invariant aspects of the affine group
representations. (ii) Once all contributions to the constraints have
been placed on the same footing, then we will apply the formalism of
\cite{CIS} to the quantization of the coupled theory and to the
construction of elements of the physical Hilbert space.

This paper proceeds as follows.  In section 2 we establish some
notations and conventions.  Then we review the main results of
\cite{LASZLO1} detailing the gravity-matter coupling for bosonic
fields in the metric representation.  The incorporation of spin 1/2
fermions will entail the introduction of a vierbein, which we
incorporate via the ADM triad-extrinsic curvature/Ashtekar variables.  The purpose of section 2 is
to establish the matter contributions to a functional $M$, whose
role will be that of the dilator element comprising the affine Lie
algebraic Hamiltonian constraint.  In Section 3 we present a
detailed analysis of the fermionic action, and the fermionic
contributions to the constraints.  A main result is that in spite of
the presence of extrinsic curvature terms in the Hamiltonian, one
still ends up with a hermitian contribution to $M$.  In section 4 we
put everything together, phrasing the Hamiltonian constraint as a
Poisson bracket between the functional $M$ and the local volume
functional $V(x)$.  Then we quantize the theory, importing the
results of \cite{CIS}, and construct elements of the physical
Hilbert space $\textbf{H}_{Phys}$ satisfying all of the constraints.
Section 5 is a summary and discussion section, including some
potential areas for future investigation.
\section{Preliminaries for the Hamiltonian constraint}

Before proceeding with the aims of this paper, let us first collect
some relevant basic results and establish some notations and
conventions.  The starting Lagrangian density for the matter fields
of this paper will be given by
\begin{eqnarray}
\label{MATTERFIELDS}
L=\frac{1}{4}\det( {^4}e)G_{\alpha\beta}F^{\alpha}_{\mu\nu}F^{\beta}_{\rho\sigma}g^{\mu\rho}g^{\nu\sigma}\nonumber\\
+\frac{1}{2}\det( {^4}e)G_{IJ}g^{\mu\nu}(D_{\mu}\phi^I)(D_{\nu}\phi^J)-\det( {^4}e)U(\psi^r,{\psi}^\dagger_{r},\phi^I)\nonumber\\
+\frac{1}{2}(\det( {^4}e)\psi^{\dagger}_r E^{\mu}_{A}\tau^{A}i{\cal
D}_{\mu}\psi^r + h.c.).
\end{eqnarray}
\noindent It will be convenient, for the purposes of canonical
decomposition into 3+1 form, to put in place some
 conventions for labeling indices.  Beginning Greek symbols $\alpha,\beta,\dots$ will denote indices
 labeling the non-gravitational Yang--Mills gauge group ${\cal G}$, which for the conventional Standard
 Model is $SU(3)\otimes{SU}(2)\otimes{U}(1)$ or its extension to the gauge groups of Grand Unified Theories, whereas symbols from the middle
 $\mu,\nu,\dots$ will refer to 4-dimensional spacetime indices.  We will use symbols from the middle
 of the Latin alphabet $i,j,k,\dots$ to denote 3-dimensional spatial indices, and from the beginning
 $a,b,c\dots$ to denote left-handed gravitational $SO(3, C)$ Lorentz indices,
 each set taking values $1,2,3$.  So the Yang--Mills gauge connection and electric field will be given by
 $A^{\alpha}_i,\widetilde{\Pi}^i_{\alpha}$; and the Ashtekar variables, a $SL(2, C)$ connection and densitized
 triad are given
by $A^a_i,\widetilde{E}^i_a$.\footnote{We will often utilize the
tilde to denote objects of density weight one.  We will omit the
tilde for the local volume element $V(x)$ to avoid cluttering the
notation.  Its density weight will be clear from the context.}
Upper case Latin letters $A, B,... $ from the beginning of the
alphabet denote $SO(3,1)$ Lorentz indices while Latin symbols from
$I,J,\dots$ will label the multiplet of Higgs fields, which
transform in a given representation of ${\cal G}$. The corresponding
phase space variables are given by $\phi^I,\widetilde{\pi}_I$.
Indices $\alpha$ will be raised by $G^{\alpha\beta}$, suitable
Killing--Cartan forms on ${\cal G}$, while indices $I$ will be
lowered by the constant fiber metric $G_{IJ}$ on the Higgs bundle.
Finally, for the phase space variables of the two-component
left-handed spin 1/2 fermions $\psi^{r},\widetilde{\pi}_{r}$,
symbols from the end of the Latin alphabet $r,s,\dots$ will
represent \emph{all} the flavor indices, which can be extended to
include all generations of fermions.  To avoid cluttering the
notation, spinorial indices shall be suppressed by using matrices
and writing each Weyl fermion as a two-component single column
matrix.
\par \indent
The Standard Model and Grand Unified Theories incorporate all
fermionic particles as left-handed multiplet(s) of Weyl fermions
transforming according to a certain representation (which we have
denoted generically by $T$) of the Yang-Mills gauge group. The total
covariant derivative for minimal coupling of fermions to gauge and
gravitational forces is ${\cal D}_i\psi^{r}= \partial_i\psi^{r}
-iA^{\alpha}_i({T}_\alpha)^r_s\psi^s -
iA_{ia}\frac{{\tau_{a}}}{2}\psi^{r}$, wherein $A_{ia}$ is precisely
the Ashtekar (anti)self-dual spin connection gauging the local
$SL(2,C)$  Lorentz symmetry \cite{Ashtekar}, $\tau^{a=1,2,3}$ denote
the standard Pauli matrices, and $\Gamma_{ia}$ denotes connection
coefficients of the intrinsic Levi-Civita connection compatible with
the triad $E^i_a$.  It will also be convenient to define the
following version of the fermionic covariant derivative
$D^\Gamma_i\psi^{r}=\partial_i\psi^{r}-i
A^{\alpha}_i({T}_\alpha)^r_s\psi^{s}-i\Gamma_{ia}\frac{\tau^a}{2}\psi^{r}$,
which subtracts out the extrinsic curvature part of ${\cal D}$. The
Yang--Mills covariant derivative acts on the Higgs multiplet with
representation ${\cal T}$ as $D_i\phi^I=\partial_i\phi^I-i
A^{\alpha}_i({\cal T}_\alpha)^I_J\phi^J$. Also of note is that the
gravitational variables are singlets under the Yang-Mills gauge
group and vice-versa. Finally, in this paper we will use the
definition $G=\frac{8\pi G_{Newton}}{c^{3}}$.
\subsection{Results from the metric-ADM formalism}

Let us now set the stage by reviewing some of the main results of
\cite{LASZLO1}, which demonstrates the coupling of bosonic matter
fields to metric-based gravity within an affine Lie-algebraic
framework.  In \cite{LASZLO1} the following integrals over 3-space
$\Sigma$ were defined 

\begin{eqnarray}
\label{DEFINED} V[n]=\int_{\Sigma}d^3x~n\sqrt{h};~~
T[f]=\frac{2}{3G}\int_{\Sigma}d^3x~f\widetilde{\pi}^{ij}h_{ij},
\end{eqnarray}
where $(h_{ij},\widetilde{\pi}^{ij})$ are the spatial 3-metric and
its conjugate momentum, and $n$, $f$ are arbitrary functions on
$\Sigma$. Essentially, $V[n]$ is Misner's, and $T[f]$ is York's
(integrated) time function, which have the following Poisson bracket

\begin{eqnarray}
\label{THEPOISSONBRACKET} \{T[f],V[n]\}=V[fn].
\end{eqnarray}
\noindent This implies that if $(\phi,\widetilde\pi)$ are the
canonical variables of \emph{any} matter field whose energy density
$\widetilde{\mu}=\sqrt{h}\mu$ does not depend on the
\emph{gravitational} ADM canonical momentum $\widetilde\pi^{ij}$,
i.e. $\mu=\mu(h_{ij},\phi,\widetilde\pi)$, then the matter part of
the Hamiltonian constraint of the Einstein--matter system can be
written as the pure Poisson bracket
\begin{eqnarray}
\label{MATTERCONSTRAINT} \Bigl\{T[\mu],V[n]\Bigr\}=\int_\Sigma d^3x
n\mu\sqrt{h}.
\end{eqnarray}
\noindent This, together with the fact that the gravitational part
of the Hamiltonian constraint can be recovered as the Poisson
bracket of the imaginary part of the Chern--Simons functional
$I_{CS}$ (built in the spinor representation from the Sen
connection) and the $V[N]$ above \cite{CIS,LASZLO,LASZLO1}, yield
that the Hamiltonian constraint of the Einstein--matter system is
the Poisson bracket of $\frac{1}{G^2}{\rm Im}\,I_{CS}+T[\mu]$ and
$V[n]$. Here the matter field may be a phenomenological matter
model, e.g. a fluid or a more general elastic material.\par \indent
In \cite{LASZLO1} it was shown explicitly that
(\ref{MATTERCONSTRAINT}) holds for any Yang--Mills--Higgs system,
where the energy densities of the Yang--Mills and Higgs fields are
(rewritten in density weight one form)
\begin{eqnarray}
\label{ENERGYDENSITIES}
&{}&\widetilde\mu_{YM}=\frac{1}{2\sqrt{h}}h_{ij}\bigl(G^{\alpha\beta}
  \widetilde{\Pi}^i_{\alpha}\widetilde{\Pi}^j_{\beta} +G_{\alpha\beta}
  \widetilde{B}^{\alpha{i}}\widetilde{B}^{\beta{j}}\bigr), \nonumber \\
&{}&\widetilde\mu_H=G^{IJ}\frac{\widetilde{\pi}_I\widetilde{\pi}_J}{2\sqrt{h}}+
  \frac{1}{2}\sqrt{h}h^{ij}G_{IJ}(D_i\phi^{I})(D_j\phi^J)+\sqrt{h}U,\nonumber\\
\end{eqnarray}
\noindent wherein
$\widetilde{B}^{i\alpha}=\frac{1}{2}\widetilde{\epsilon}^{ijk}F^{\alpha}_{jk}$
denotes the magnetic field strength of $A^\alpha_i$.  Here
$U=U(\phi^I)$ is a scalar potential which can include the mass and
self-interaction terms for the Higgs field, but is independent of
the momenta of gravitational variables.  Note that none of the terms
in (\ref{ENERGYDENSITIES}) depends on the extrinsic curvature of
$\Sigma$, i.e. on the ADM canonical momentum $\widetilde\pi^{ij}$.
\par \indent We will extend these results to the fundamental fields
of the Standard Model, which entails generalization of the
gravity-matter coupling to include spin 1/2 fermions.  Since a spin
1/2 fermion constitutes a finite dimensional representation of the
Lorentz group and not of the group of general coordinate
transformations, a vierbein is needed in the description of
fermionic coupling to gravity.  An expression for
$\widetilde{\mu}_F$ can be derived from the 3+1 form of the standard
Lagrangian for the Weyl spinor fields via the standard canonical
formalism, or simply by rewriting the time-time component of the
energy-momentum tensor for the Weyl spinor field (see, for instance,
Ref.\cite{PR}) by identifying $\widetilde\pi_r =-i\sqrt{h}\psi
^\dagger_r(n_A\tau^{A})$ as the momentum conjugate to the
two-component Weyl spinor $\psi^r$ ($\widetilde\pi_r$ is a
two-component row vector in matrix notation if we write $\psi^r$ as
a two-component column vector). Here $n_{A}\tau^A$ denotes the
spinorial form of the future pointing unit timelike normal to
(3-space) $\Sigma$ in the spacetime. The derivation of the fermionic
contribution to the Hamiltonian constraint is carried out in Section
3, with the result
\begin{eqnarray}
\label{ENERGYDENSITIES1}
&{}&\widetilde\mu_F=-\frac{1}{2\sqrt{h}}\bigl(\widetilde\pi_r{\widetilde
E}^i_a\tau^a
 {\cal D}_i\psi^{r}+c.c.\bigr).
\end{eqnarray}
\noindent Once fermions have been introduced into the theory, the
potential can be extended
$U\rightarrow{U}(\phi^I,\psi^{r},\widetilde{\pi}_{s})$ to include
Higgs-fermion mass couplings, still remaining a pure algebraic
expression of its arguments independent of $\widetilde{\pi}^{ij}$.

\subsection{Inclusion of fermions, and Ashtekar-ADM triad variables}

We will now proceed with the incorporation of fermions into the
theory.  Since a vierbein will be needed, we will transition
directly into a nonmetric formulation of GR, where the triad
replaces the metric as a fundamental degree of freedom.  Since the
fermionic energy density (\ref{ENERGYDENSITIES1}) depends on the
extrinsic curvature, this poses an additional challenge in relation
to its bosonic counterparts in writing the analogue of
(\ref{MATTERCONSTRAINT}), which will need to be dealt with. Also, if
one were to naively resort to the Ashtekar self-dual formalism in its
original form, one would be confronted with issues of reality
conditions unique to the existence of fermions.  Both difficulties can be circumvented.
The correspondence to an affine group structure is precise only if the generators are hermitian. As discussed in \cite{CIS}, for pure gravity this selects the imaginary part, $Q$, of the Chern-Simons functional rather the full Chern-Simons expression functional of the Ashtekar connection. Similarly, the corresponding hermitian part of the fermionic Hamiltonian density will be selected in the final expression of $M$ in (\ref{GEE}), and the affine algebra derived is equivalent to the usual real total Hamiltonian constraint expressed in terms of the densitized triad variable and its conjugate momentum.

\par \indent
The fundamental extrinsic curvature and densitized triad ADM
variables $(\widetilde{E}^{ia}, K_{ia})$ have non-trivial Poisson
bracket $\{\widetilde{E}^{ia}(x), K_{jb}(y)\}_{P.B.}=G
\delta_{b}^{a}\delta_{j}^{i}\delta^{(3)}(x,y)$.  To express
(\ref{ENERGYDENSITIES}) in terms of these variables, the pertinent
entities in (\ref{ENERGYDENSITIES}) can be expressed as
\begin{eqnarray}
\label{HAM2}
&h=\det(h_{ij})=\det{\widetilde E}=\frac{1}{3!}\eps\epsilon_{abc}{\widetilde E}^{ia}{\widetilde E}^{jb}{\widetilde E}^{kc} , \nonumber\\
&h^{ij} = \frac{\widetilde{E}^i_a
\widetilde{E}^j_a}{\det\widetilde{E}}, \quad h_{ij} =\I \J
det\widetilde{E},
\end{eqnarray}
where $\I$ and $\J$ are defined as the weight $-1$ inverse
densitized triads and the imaginary part of the Chern-Simons
functional of the Ashtekar connection, $Q$, can be expressed totally
in terms of $\Gamma_{ia}$ (which depends only on the triad $e^a_i$)
and the extrinsic curvature $K_{ia}$ when the substitution for
$A_{ia}= i{K}_{ia} + \Gamma_{ia}$ is carried out in $Q$. The
explicit expression is detailed in Appendix B of \cite{CIS}.\par
\indent The density weight one version of the Hamiltonian constraint
for gravity with cosmological constant $\Lambda$, with Higgs and
spin $\frac{1}{2}$  Weyl fermionic multiplets minimally coupled to
Yang-Mills fields is given by

\begin{eqnarray}
\label{HAM}
\widetilde{H}=\widetilde{H}^{grav}+\sqrt{h}\bigl(\mu_H+\mu_{YM}
+ \mu_{F}\bigr)=0,
\end{eqnarray}
wherein the respective Hamiltonian densities (the density weight
zero form in the case of the matter
 contributions) are\footnote{The matter Hamiltonian densities are of mass dimension $[\mu]=4$, hence the factor of $\frac{1}{G^2}$ which multiplies the dimensionless $Q$.  This balances the mass dimensions of the gravitational with matter terms in accordance with the Einstein equations.}

\begin{eqnarray}
\label{HAM1}
\widetilde{H}^{grav}&=&\frac{1}{G^2}\bigl(\{Q,V(x)\}+\frac{G\Lambda}2V(x)\bigr);\nonumber\\
\mu_H&=&G^{IJ}\frac{\widetilde{\pi}_I\widetilde{\pi}_J}{2(\hbox{det}\widetilde{E})}+\frac{\widetilde{E}^i_a\widetilde{E}^j_a}{2(\hbox{det}\widetilde{E})}G_{IJ}(D_i\phi^{I})
(D_j\phi^J)+U(\phi^I, \psi^{r}, \widetilde{\pi}_{s});\nonumber\\
\mu_{YM}&=&\frac{1}{2}\I \J\bigl(G^{\alpha\beta}\widetilde{\Pi}^i_{\alpha}\widetilde{\Pi}^j_{\beta} +G_{\alpha\beta}\widetilde{B}^{\alpha{i}}\widetilde{B}^{\beta{j}});\nonumber\\
\mu_{F}&=&-\frac{1}{h}\widetilde{\pi}_r{\widetilde
E}^{i}_{a}\frac{\tau^a}{2}\bigl(\partial_{i}\psi^r +( K_{ib}
-{i}\Gamma_{ib})\frac{\tau^b}{2}\psi^r
-iA^{\alpha}_i(T_{\alpha})^r_s\psi^s\bigr) + h.c. \cr
&=&-\{\frac{1}{h}\widetilde{\pi}_r\widetilde{E}^{i}_{a}\frac{\tau^{a}}{2}D^{\Gamma}_{i}\psi^r+h.c.\}
-\frac{i}{h}\widetilde{\pi}_r\epsilon^{abc}\widetilde{E}^{i}_{a}K_{ib}\frac{\tau^{c}}{2}\psi^r.
\end{eqnarray}

In the above, $Q$ is the imaginary part of the Chern--Simons
functional of the Ashtekar connection $A^a_i$; and the expression of
$\widetilde{H}^{grav}$ in this affine Lie-algebraic form is a key
result of \cite{CIS}. The fermionic Hamiltonian density
$\widetilde{\mu}_{F}$ can be obtained from the {\it hermitian}
action of fermions coupled to gravitational and Yang-Mills
connections\cite{Thiemann3}. The resultant fermionic contribution to
the Hamiltonian constraint indicated by
$\widetilde{\mu}_{F}=\sqrt{h}\mu_F$ with $\mu_F$ as in (\ref{HAM1})
can be obtained by substituting the Ashtekar connection $A_{ia} = i
K_{ia}
 +\Gamma_{ia}$ into the hermitian fermionic action\cite{Thiemann3}.\footnote{The derivation is provided in the next section.} This accounts for why $D^\Gamma_i\psi$ in the final line of (\ref{HAM1}) contains $\Gamma_{ia}$ but not $A_{ia}$.\par
\indent It will also be useful to have the density weight one
functional $T[f]$ in Ashtekar variables, given by the densitized
trace of the extrinsic curvature
\begin{eqnarray}
\label{ONEHALF4}
T[1]=\frac{2}{3G}\int_{\Sigma}d^3xK^a_i\widetilde{E}^i_a=\frac{2}{3iG}\int_{\Sigma}d^3x
(A^a_i-\Gamma^a_i)\widetilde{E}^i_a.
\end{eqnarray}

The main result of \cite{LASZLO1} is that, including the bosonic
matter with gravitational contributions, the local Hamiltonian
constraint can be written as the Poisson bracket $\{M,V(x)\}$ for a
suitably chosen functional $M$, in direct analogy to the first
equation of (\ref{HAM1}) absent the cosmological constant
term.\footnote{It is crucial that the functional $M$ be chosen so as
to produce the density weight one form of the Hamiltonian
constraint. $M$ is thus diffeomorphism invariant. Moreover, as noted
in \cite{CIS,Thiemann3}, only for this density weight does one have
a constraint with good self-regularizing properties, which is free
of ultraviolet divergences.}  The present work aims to draw
attention to the fact that the Hamiltonian constraint of the
\emph{complete} Standard Model or even Grand Unified Theories (GUTs)
coupled to gravitation can be expressed in the form of an affine
algebra with Hermitian generators.  This, combined with the ability
to construct coherent states using gauge invariant, diffeomorphism
invariant group elements of the affine group, leads directly to the
existence of elements of the physical Hilbert space
$\textbf{H}_{Phys}$ of the coupled theory.  These states thus form
unitary, irreducible representations of the affine group $Aff(R)$.  Moreover, the
states implement at a quantum level the positivity of the spectrum of the 
volume operator.

\section{Fermionic contribution to the constraints}

In this section, we derive the fermionic contribution to the
super-Hamiltonian and kinematic constraints from the
\emph{hermitian} action of fermions coupled to gravitation and
Yang--Mills theory; and determine the functional whose Poisson
bracket with the volume element reproduces the former. To wit, the
full multiplet of two-component left-handed Weyl fields,
$\psi^r$, coupled to gravity and unified
Yang--Mills theory has the Hermitian action,
\begin{eqnarray}
\label{WEYLACTION} S_{Weyl}&=&\frac{1}{2}\int d^{4}x(\det
{^4}e)\psi_r^\dagger E^{\mu}_{A}\tau^{A}i\mathcal{D}_{\mu}\psi^r +
h.c.\cr &=&\frac{1}{2}\int dt\int d^{3}x(N
e)\psi_r^\dagger(\mathcal{E}^{\mu}_{A}-n_{A}n^{\mu})\tau^{A}i\mathcal{D}_{\mu}\psi^r+
h.c. \cr &=&\frac{1}{2}\big[\int dt\int d^{3}x
Ne\psi_r^\dagger\mathcal{E}^{\mu}_{A}\tau^{A}i\mathcal{D}_{\mu}\psi^r
-\int dt\int d^{3}x e\psi_r^\dagger (n_{A}\tau^A)
t^{\mu}i\mathcal{D}_{\mu}\psi^r\cr &&+\int dt\int d^{3}x N^\mu
e\psi_r^\dagger( n_{A}\tau^{A})i\mathcal{D}_{\mu}\psi^r \big]+ h.c.
\end{eqnarray}
In Eq.(\ref{WEYLACTION}), the definition for $\tau^{A = 0,1,2,3}$
is $(\tau^0 = I_2, \tau^{a=1,2,3}$ = Pauli matrices), and
$\psi_r^\dagger ={\bar\psi}_r^T$ (the overline denotes complex
conjugation). We first decomposed the vierbein as $E^\mu_A =
(\mathcal{E}^{\mu}_{A}-n_{A}n^{\mu})$, with the normal vector
$n^\mu$ satisfying $n_\mu n^\mu = -1$ and $n_A \equiv n_\mu
E^\mu_A$.
 In terms of the lapse and shift functions, the normal can be expressed as $n^\mu = \frac{(t^\mu-N^\mu)}{N},$ where $n_{\mu}N^{\mu}=0$. The fermion conjugate momentum is $\tilde{\pi}_r  =\frac{\partial {L}}{(\partial\pounds_t\psi^r)} = -ie\psi_r^\dagger(n_A\tau^A)$,  wherein $\pounds_t$ is the Lie derivative with respect to the vector field $t^\mu\partial_\mu = (Nn^\mu + N^\mu)\partial_\mu$.
 Under the action of the Lorentz group, $\widetilde{\pi}_r$  transforms as $\widetilde{\pi}^{\prime}_r= \widetilde{\pi}_r u^{-1}_L$  i.e. inversely as the left-handed spinor ${\psi'}^r =u_L\psi^r \quad \forall\, u_L\in SL(2,C)$.  Carrying out a Legendre transformation into the Hamiltonian description, we can read off the spin 1/2 contributions to the intial value constraints.

\subsection{The Hamiltonian constraint}

So the contribution to the Hamiltonian constraint, which has density
weight one and $N$ as Lagrange multiplier, is thus
\begin{eqnarray}
\label{hd} \widetilde{\mu}_{F}=\sqrt{h}\mu_{F}&=&
-\frac{1}{2}e\psi_r^\dagger\mathcal{E}^{\mu}_{A}\tau^{A}i\mathcal{D}_{\mu}\psi^r\
+ h.c. \cr &=& -\frac{1}{2}e\psi_r^\dagger
(n_B\tau^B)(n_C{\bar\tau}^C)\mathcal{E}^{\mu}_{A}\tau^{A}i\mathcal{D}_{\mu}\psi^r\
+ h.c. \cr
&=&\frac{1}{2}\widetilde{\pi}_r(n_C\mathcal{E}^{\mu}_{A}){\bar\tau}^C\tau^{A}\mathcal{D}_{\mu}\psi^r+
h.c.
\end{eqnarray}
In the second line of Eq.(\ref{hd}), we have inserted the identity
matrix $(n_B\tau^B)(n_C{\bar\tau}^C)=I_2$, with ${\bar\tau}^A
\equiv (I_2, - \tau^{1,2,3})$. Furthermore, the entity
$(n_C\mathcal{E}^{\mu}_{A}){\bar\tau}^C\tau^{A}$  transforms
according to the adjoint representation of $SL(2,C)$. It is spatial
since $n_\mu(n_C\mathcal{E}^{\mu}_{A}){\bar\tau}^C\tau^{A} =0$, and
also traceless. From $n_An^A = n_\mu n^\mu  =-1$, the choice of $n_A
= (-1, 0,0,0)$ yields
$(n_C\mathcal{E}^{\mu}_{A}){\bar\tau}^C\tau^{A}=n_C (E^\mu_A +
n_{A}E^\mu_B n^{B}){\bar\tau}^C\tau^{A} = -E^\mu_a\tau^a$. Since
$n_\mu(E^\mu_a\tau^a)=0$, a further choice of $n_\mu =
-N\delta_\mu^0$  renders $E^0_a =0$. Thus, the fermionic
contribution to the Hamiltonian constraint reduces to
\begin{eqnarray}
\widetilde{\mu}_{F}
&=&-\widetilde{\pi}_rE^{i}_{a}\frac{\tau^a}{2}\mathcal{D}_{i}\psi^r
+ h.c. \cr \label{hdd}
&=&-\widetilde{\pi}_rE^{i}_{a}\frac{\tau^a}{2}\bigl(\partial_{i}\psi^r
+\frac{1}{4}A_{iAB}{\bar\tau}^A\tau^B\psi^r-iA^{\alpha}_i(T_{\alpha})^r_s\psi^s\bigr)
+ h.c.\cr
&=&-\widetilde{\pi}_rE^{i}_{a}\frac{\tau^a}{2}\bigl(\partial_{i}\psi^r
+( A_{i0b}
+\frac{i}{2}\epsilon_{0bcd}A^{cd}_i)\frac{\tau^b}{2}\psi^r-iA^{\alpha}_i(T_{\alpha})^r_s\psi^s\bigr)
+ h.c.\cr &=&-\frac{1}{\sqrt h}\widetilde{\pi}_r{\widetilde
E}^{i}_{a}\frac{\tau^a}{2}\bigl(\partial_{i}\psi^r +( K_{ib}
-{i}\Gamma_{ib})\frac{\tau^b}{2}\psi^r-iA^{\alpha}_i(T_{\alpha})^r_s\psi^s\bigr)
+ h.c.
\end{eqnarray}
In the final step, we have denoted the spin connection $A_{iAB}$ by
$K_{ia}$ = $A_{i0a}$ and $\Gamma_{ia}
=\frac{1}{2}\epsilon^{0abc}A_{ibc}$. In the absence of fermionic
matter, $e^a_{(i}K_{j)a}$ is the extrinsic curvature. In general,
$K_{ia}$ is the conjugate momentum to ${\widetilde E}^{ia}$ while
$\Gamma_{ia}$ is the torsionless connection compatible with the
triad.

We would like to construct a functional $M_{1/2}$ on the classical
phase space such that
\begin{eqnarray*}
\{M_{1/2}, V(x)\}_{P.B.}=\widetilde{\mu}_{F}(x);
\end{eqnarray*}
wherein the volume element is given by
$V(x)=\sqrt{\frac{1}{3!}\eps\epsilon^{abc}\widetilde{E}^{i}_{a}\widetilde{E}^{j}_{b}\widetilde{E}^{k}_{c}}=\sqrt{\det({\widetilde
E})}$.  To wit, the following identities are useful
\begin{eqnarray}
 &\{V(x), K_{ia}(y)\} = \frac{G}{2}e_{ia}(x)\delta^{(3)}(x,y), \nonumber\\
&\{V(x), \int_{\Sigma}  K_{ia}{\widetilde E}^{ia} d^3y \}=
\frac{3G}{2}V(x).
\end{eqnarray}
Due to the hermitization, the term containing $K_{ia}$ in
(\ref{hdd}) is just $-i\epsilon^{abc}{\widetilde E}^i_aK_{ib}
(\widetilde{\pi}_r\tau_{c}\psi^r)$ which has vanishing Poisson
bracket with $V(x)$. It follows that ${\widetilde\mu}_F$ commutes with
$V(x)$; and a solution for $M_{1/2}$ is
\begin{eqnarray}
\label{2} M_{1/2} =T[-\mu_F]=\int d^{3}y
\Bigl(\frac{2\widetilde{K}}{3G\det({\widetilde
E})}\Bigr)\{\widetilde{\pi}_r\widetilde{E}^{i}_{a}\frac{\tau^{a}}{2}\bigl(\partial_{i}\psi^r
+(K_{ib}-i\Gamma_{ib})\frac{\tau^{b}}{2}\psi^r\nonumber\\
-iA^{\alpha}_i(T_{\alpha})^r_s\psi^s\bigr)+ h.c. \};
\end{eqnarray}
wherein ${\tilde K} \equiv K_{ia}{\tilde E}^{ia}$, and
$\frac{2}{3G}\int {\tilde K}d^3y = T[1]$ from (\ref{ONEHALF4}). Note
that the integrand in $M_{1/2}$ is of density weight one and thus
$M_{1/2}$ is diffeomorphism-invariant, and it is moreover
classically real since we started from the Hermitian Weyl action
(\ref{WEYLACTION}).

Similarly, the solutions for the Higgs and Yang--Mills contributions
are
\begin{eqnarray*}
\{M_{0}, V(x)\}_{P.B.}=\widetilde{\mu}_{H}(x), \quad \{M_{1},
V(x)\}_{P.B.}=\widetilde{\mu}_{YM}(x),
\end{eqnarray*}
and
\begin{eqnarray*}
M_{0}=T[-\mu_H], \quad
M_{1}=T[-\mu_{YM}].
\end{eqnarray*}

Replacing $iK_{ia} + \Gamma_{ia}$ by $A_{ia}$  achieves the
alternative to express\footnote{The subscripts on $M$ refer to the
spins of the respective contributions to $M$.} $M_{0}, M_{1/2},
M_{1}$ in terms of Ashtekar variables $(\widetilde{E}^{ia}, A_{ia})$; but it
should be noted that unlike Ashtekar's totally chiral
prescription\cite{Ashtekar}, the hermitian conjugate operation in
the action and Hamiltonian involves the spin connection of the
opposite chirality. Thus spin connections of both chiralities $\pm
iK_{ia} + \Gamma_{ia}$ are needed to render $M_{1/2}$ real.

\subsection{The kinematic constraints}

We have written the fermionic contribution to the Hamiltonian
constraint as a Poisson bracket involving the functional $M_{1/2}$.
We will next derive the spin 1/2 contributions to the kinematic
constraints. From the third term on the right hand side of
(\ref{WEYLACTION}), one has
\begin{eqnarray}
\label{VECT}
-\frac{1}{2}\int{dt}\int{d}^3xN^{\mu}e\psi_r^{\dagger}(n_A\tau^A)i\mathcal{D}_{\mu}\psi^r+h.c.
=\frac{1}{2}\int{dt}\int{d}^3x\widetilde{\pi}_rN^{\mu}\mathcal{D}_{\mu}\psi^r+h.c.
\end{eqnarray}
\noindent Stipulating that $n_{\mu}N^{\mu}=0$, then in the chosen gauge the shift vector
is purely spatial and one can make the identification
$N^{\mu}\rightarrow{N}^i$, which implies the fermionic contribution to
the smeared vector constraint
\begin{eqnarray}
\label{VECT1}
H_i^{Weyl}[N^i]=\frac{1}{2}\int{d}^3xN^i\widetilde{\pi}_r\mathcal{D}_i\psi^r+h.c.
\end{eqnarray}
\noindent The second term of (\ref{WEYLACTION}) yields
\begin{eqnarray}
\label{VECT2}
\frac{1}{8}\int{dt}\int{d}^3xe\psi_r^{\dagger}(n_A\tau^A)it^{\mu}A_{\mu{AB}}\overline{\tau}^A\tau^B\psi^r+h.c.\nonumber\\
=-\frac{1}{2}\int{dt}\int{d}^3x\widetilde{\pi}_r(\frac{\tau^c}{2})\psi^r{t}^{\mu}\bigl(A_{\mu{0}c}+\frac{i}{2}\epsilon_{0cab}A^{ab}_{\mu}\bigr)+h.c.
\end{eqnarray}
\noindent
where
$(-i)t^{\mu}\bigl(A_{\mu{0}c}+\frac{i}{2}\epsilon_{0cab}A^{ab}_{\mu}\bigr)$ is a Lagrange multiplier.  So the fermionic contribution to the smeared gravitational
Gauss' law constraint is
\begin{eqnarray}
\label{VECT3}
G_c^{Weyl}[\eta^c]=-\int{d}^3x\eta^c[i\widetilde{\pi}_r(\frac{\tau^c}{2})\psi^r]
\end{eqnarray}
\noindent
for some complex SO(3,C)-valued parameter $\eta^c(x)$.  Similarly, one can read off the analogous Yang--Mills part
\begin{eqnarray}
\label{VECT4}
\frac{1}{2}\int{dt}\int{d}^3xe\psi_r^{\dagger}(n_A\tau^A)it^{\mu}\bigl(-iA^{\alpha}_{\mu}(T_{\alpha})^r_s\psi^s\bigr)+h.c.
=\frac{i}{2}\int{dt}\int{d}^3xt^{\mu}A^{\alpha}_{\mu}\widetilde{\pi}_r(T_{\alpha})^r_s\psi^s+h.c.,
\end{eqnarray}
\noindent
where ${t}^{\mu}A^{\alpha}_{\mu}$ is the corresponding Lagrange multiplier.  So we can extract the fermionic contribution to the smeared Yang--Mills Gauss' law constraint, for some $\cal{G}$-valued real parameter $\eta^{\alpha}(x)$, given by
\begin{eqnarray}
\label{VECT5}
G_{\alpha}^{Weyl}[\eta^{\alpha}]=\int{d}^3x\eta^{\alpha}[\widetilde{\pi}_r(iT_{\alpha})^r_s\psi^s].
\end{eqnarray}
\noindent \par
\indent In gravitational theories incorporating local
gauge-invariance, the gauge-invariant vector (super-momentum)
constraint, with the shift function as Lagrange multiplier,
generates gauge-covariant Lie derivatives rather than pure spatial
Lie derivatives. The constraint itself is gauge-invariant under all
local gauge symmetries of the theory. By adding suitable
combinations of the Gauss' Law constraints generating local gauge
symmetries, a non-gauge-invariant corresponding constraint can be constructed
to generate true spatial Lie derivatives of the dynamical variables.
It follows that any spatial
integral over the entire Cauchy surface of any tensor density weight
one gauge-invariant composite of dynamical variables will be
invariant  under both spatial diffeomorphisms and also under local
gauge symmetries generated by the Gauss' Law constraints. Thus the
$M^{Grav}, M_0, M_1, M_{1/2}$ in Eq. (\ref{GEE}) entities that we
construct later on will all be gauge and spatial diffeomorphism
invariant as they are integrals of weight one tensor densities of
gauge-invariant composites of dynamical variables.

\section{Putting it all together}

We have addressed all of the ingredients necessary to construct a
theory which respects the Hamiltonian constraint as well as the
kinematic constraints for gravity coupled with the matter fields of
the Standard Model in the Ashtekar-ADM triad variables.  So we may now proceed
with a view toward quantization of the theory and the construction
of elements of the physical Hilbert space $\textbf{H}_{Phys}$.
Putting everything together, recall that
$\widetilde{H}=\widetilde{H}^{grav}+\widetilde{\mu}_H+\widetilde{\mu}_F+\widetilde{\mu}_{YM}$.\footnote{where
$\widetilde{H}^{grav}=\frac{1}{G^2}\bigl(\{Q,V(x)\}+\frac{G\Lambda}{2}V(x)\bigr)$,
$\{M_0,V(x)\}=\widetilde{\mu}_H$,
$\{M_{1/2},V(x)\}=\widetilde{\mu}_F$, and
$\{M_1,V(x)\}=\widetilde{\mu}_{YM}$.} We can define the following
contributions to the functional
$M=M^{Grav}+G^2\bigl(M_0+M_{1/2}+M_1\bigr)$ with
\begin{eqnarray}
\label{GEE}
&&M^{Grav}=Q,\nonumber\\
&&M_0=T[-\mu_H],\nonumber\\
&&M_1=T[-\mu_{YM}],\nonumber\\
 &&M_{1/2}=\int d^{3}y
[\frac{2\widetilde{K}}{3G\det({\tilde
E})}]\{\widetilde{\pi}_r\widetilde{E}^{i}_{a}\frac{\tau^{a}}{2}{\cal
D}_i\psi^r+ h.c. \}=T[-\mu_F]
\end{eqnarray}
with $T[f(x)]\equiv\frac{2}{3G}\int d^3x f K_{ai}{\tilde E}^{ai}$ and
$\widetilde{K}=K^a_i\widetilde{E}^i_a$.

Then the Hamiltonian constraint of the Standard Model coupled to
gravity can be written as a Poisson bracket
\begin{eqnarray}
\label{GEE1} \{M,V(x)\}+\frac{G\Lambda}{2}V(x)=0.
\end{eqnarray}
\noindent The \emph{full} Hamiltonian constraint for gravity coupled
with matter fields, in direct analogy to \cite{CIS}, can be written
as an affine Lie algebra.  Since all terms of $M$ are real, then
their quantum versions will correspond to Hermitian operators.
 Upon making the replacement $Q\rightarrow{M}$ in \cite{CIS} and
carrying out the steps for quantization, then the Hamiltonian
constraint becomes the affine Lie algebra
\begin{eqnarray}
\label{GEE2}
\widehat{H}(x)\vert\psi\rangle=[i\widehat{M},\widehat{V}(x)]\vert\psi\rangle+\lambda\widehat{V}(x)\vert\psi\rangle=0;~~\lambda=\frac{\hbar{G}\Lambda}{2}.
\end{eqnarray}
\noindent So the solution to the quantum Hamiltonian constraint
(\ref{GEE2}) is tantamount to the statement that $\vert\psi\rangle$
must form a representation space of the affine group $Aff(R)$.  But
for $\vert\psi\rangle$ to be a physical state, it must additionally
be gauge invariant\footnote{That is, gauge invariant with respect to
the gravitational and the nongravitational Yang--Mills gauge
groups.} and diffeomorphism invariant
\begin{eqnarray}
\label{INVARIANT}
\widehat{H}_i(x)\vert\psi\rangle=\widehat{G}_a(x)\vert\psi\rangle=\widehat{G}_{\alpha}(x)\vert\psi\rangle=0.
\end{eqnarray}
\noindent Let
$\vert\eta\rangle=\vert{0},0\rangle\in{Ker}\widehat{H}(x)$ be a
gauge invariant, diffeomorphism invariant and normalized fiducial
state.  We would like to construct gauge invariant,
diffeomorphism-invariant states
$\vert{a},b\rangle\in{Ker}\widehat{H}_i(x),\widehat{G}_a(x),\widehat{G}_{\alpha}(x)$ as
elements of the physical Hilbert space $\textbf{H}_{Phys}$ from
affine group elements, for spatially homogeneous quantities $a$ and
$b$ of mass dimensions $[a]=0$ and $[b]=3$, given by
\begin{eqnarray}
\label{GEE3}
\vert{a},b\rangle=e^{-ia\widehat{M}}e^{-ib\widehat{V}}\vert{0},0\rangle\in\textbf{H}_{Phys}.
\end{eqnarray}
\noindent But for this to be the case, the generators $\widehat{M}$
and $\widehat{V}$ must be gauge-diffeomorphism invariant.  The
3-space volume operator $\widehat{V}$ is already
gauge-diffeomorphism invariant since it is the integral of a
geometric object of density weight one
$\int_{\Sigma}d^3x\widehat{V}(x)$, whose internal indices are
completely contracted.  For the object $\widehat{M}$, it suffices to
note via density weight counting, that the integrand of
$\widehat{M}$ is an object of density weight one, hence
$[\widehat{M},\widehat{H}_i]=0$ and $M$ is diffeomorphism invariant.
Indeed it is the convolution of $T$, which is of density weight one on
account of the densitized triad $\widetilde{E}^i_a$, and the various
contributions to $\mu$ which are of density weight zero.
Additionally, $M$ is gauge invariant (e.g.
$[\widehat{M},\widehat{G}_a]=[\widehat{M},\widehat{G}_{\alpha}]=0$)
since it is constructed from tensorial geometric objects whose gauge
indices (gravitational and nongravitational) are completely
contracted.  The result is that
$\vert{a},b\rangle\in\textbf{H}_{Phys}$ constitute a set of physical
states for the full theory of the Standard Model coupled to gravity,
through unitary irreducible representations of the affine group, in
direct analogy with the results of \cite{CIS}.\par
\indent
The affine algebra for gravity coupled with the Standard Model implies a fundamental uncertainty
relation
\begin{eqnarray}
\label{RELATION}
\frac{\Delta{V}}{\langle{V}\rangle}\Delta
{M}\geq 2\pi \Lambda L^2_{Planck},
\end{eqnarray}
\noindent
wherein $V$ is the total volume of the universe.  This is an extension of the uncertainty relation
between $Q$ and $V$ obtained in \cite{CIS}, which depends on the cosmological constant; and equation
(\ref{RELATION}) is a result of treating gravity and matter on equal quantum footing.
It is intriguing that current observations place $\Lambda L^2_{Planck} \sim 10^{-120}$ which renders the uncertainty small, but nevertheless non-vanishing, while the precise correspondence of the Hamiltonian constraint to an affine algebra is predicated upon a non-vanishing cosmological constant. Affine algebraic quantization of both gravitation and matter on equal footing may yield a new perspective on the cosmological constant problem.

\section{Summary and future research}

This work demonstrates that a complete description of the
interaction of matter and all forces  (gravitational and
non-gravitational)  can in fact be realized within a quantum affine
algebraic framework.  It has been shown that the generalization of the
affine group formalism for four dimensional, Lorentzian signature
quantum gravity in the Ashtekar variables in \cite{CIS}
straightforwardly extends to the coupling to non-gravitational
fields\footnote{Generalization to supersymmetric extensions of the
Standard Model should be straightforward via the techniques of this
paper, and application to supergravity and GUTs can as well be
examined.}.  We have utilized this framework to construct elements
of the physical Hilbert space of the Standard Model coupled to
gravity.

The construction of a physical Hilbert space for gravity has until
now remained an unsolved problem in theoretical physics, and the
main contribution of this paper has been to provide a glimpse of
certain elements of this space.  The phenomenological implications
of this result opens up various areas for further investigation, not
limited to and including:
(i) To check for consistency with symmetry breaking via the Higgs
mechanism, using the affine coherent states $\vert{a},b\rangle$ as
the starting point.  (ii) To examine the quantum gravitational
effects encoded in and induced from the states $\vert{a},b\rangle$
in the semiclassical limit below the Planck scale, and to compare
its predictions with those of quantum field theory on curved
spacetime.  It is conceivable that in this limit, imprints of the
Planck scale physics could in some sense be probed via this
formalism.



\section{Acknowledgements}
This work has been supported in part by the Office of Naval Research
under Grant No. N-000-1412-WX-30191, the National Science Council of
Taiwan under Grant No. NSC 101-2112-M-006 -007 -MY3, and the
National Center for Theoretical Sciences, Taiwan.


\begin{thebibliography}{99}
\bibitem{CIS} {Chou Ching-Yi, Eyo Ita and Chopin Soo, `Affine group
   representation formalism for four dimensional, Lorentzian, quantum
   gravity' Class. Quantum Grav. {\bf 30} (2013): 065013}

\bibitem{LASZLO} {L\'aszl\'o B. Szabados, `On the role of conformal three
   geometries in the dynamics of General Relativity', Class. Quantum Grav.
   {\bf 19} (2002) 2375-2392}

\bibitem{LASZLO1} {L\'aszl\'o B. Szabados, `A note on the Hamiltonian
   constraint in canonical GR', Class. Quantum Grav. {\bf 25} (2008): 095005}

\bibitem{PR} {R. Penrose, W. Rindler, `Spinors and spacetime', vol.2
(Cambridge University Press, 1986)}

\bibitem{Ashtekar}A. Ashtekar, `Lectures on non-perturbative canonical gravity' (World Scientific, 1991)

\bibitem{Thiemann3} {T. Thiemann, `Modern Canonical Quantum General Relativity' (Cambridge University Press, 2007)}

\bibitem{ASH1}A. Ashtekar, J,Romano, and R. Tate `New variables for gravity: Inclusion of matter', Phys. Rev. D40, 2572-2587 (1989)

\end{thebibliography}
\end{document}